\documentclass{osa-article}

\journal{oe}

\begin{document}

\title{Simulation technique of quantum optical emission process from multiple two-level atoms based on classical numerical method}

\author{Hideaki Taniyama,\authormark{1,2,*} Hisashi Sumikura,\authormark{1,2} and Masaya Notomi\authormark{1,2,3}}

\address{\authormark[1]NTT Nanophotonics Center, NTT Corporation, 3-1 Morinosato Wakamiya, Atsugi, Kanagawa, 243-0198, JAPAN\\
\authormark[2]NTT Basic Research Laboratories, NTT Corporation, 3-1 Morinosato Wakamiya, Atsugi, Kanagawa, 243-0198, JAPAN\\
\authormark[3]Department of Physics, Tokyo Institute of Technology, H-55, Ookayama 2-12-1, Meguro 152-8550, JAPAN}
\email{\authormark[*]hideaki.taniyama.fy@hco.ntt.co.jp}

\begin{abstract}
In this paper, we report a numerical method for analyzing optical radiation from a two-level atom.
The proposed method can consistently consider the optical emission and absorption process of an atom,
and also the interaction between atoms through their interaction with a radiation field.
The numerical model is based on a damping oscillator description of a dipole current, which is a classical model of atomic transition and is implemented with a finite-difference time-domain method.
Using the method, we successfully simulate the spontaneous emission phenomena in a vacuum, where the interaction between an atom and a radiated field plays an important role.
We also simulate the radiation from an atom embedded in a photonic crystal (PhC) cavity.
As a result, an atom-cavity field interaction is sucessfuly incorporated in the simulation, and the enhancement of the optical emission rate of an excited atom is explained.
The method considers the effect of the interaction between atoms through the radiated field.
We simulate the optical emission process of the multiple atoms and show that an enhancement of the emission rate can occur owing to the an atom-atom interaction (superradiance)(R. H. Dicke, Phys. Rev. {\bf 93}, 99[1954]).
We also show that the emission rate is suppressed by the effect of the destructive dipole-dipole interaction under an out-of-phase excitation condition (subradiance).
\end{abstract}

\section{Introduction}
The spontaneous emission rate is a fundamental physical quantity that governs light-matter interaction.
It is well known that the radiation process of an atom depends on its environment\cite{purcell46}.
This means that the radiation process can be controlled artificially by changing the environment.
Many research studies have been carried out aiming at the control of the radiation process in combination with a micro optical resonator, such as PhC cavities\cite{yoshie04} and whispering-gallery cavities\cite{armani03}.

The finite-difference time-domain (FDTD) method solving Maxwell's equation has become a powerful tool in computational electrodynamics.
The method has been applied to studies of the optical process in actual photonic nanostructures.
Recently, an FDTD method has been applied to phenomena including the spontaneous emission process of an excited two-level atom located in an optical cavity, which is called cavity quantum electrodynamics (cQED)\cite{vuckovic99,xu00}.
However, their numerical model adopts dipole current decay in time at a given a priori rate.
In their calculation, the decay rate was chosen to match the spontaneous emission rate of an atom in a vacuum.
Due to their formulation, which is based on a dipole model assuming attenuation beforehand, their application is limited.
It cannot be applied to pure spontaneous emission phenomenon in a vacuum.

Recently, we reported the numerical technique based on FDTD method to analyze optical process\cite{Taniyama11}.
In the report, we successfully simulated an emission process in an optical cavity.
However, the numerical model used in the calculation assumed a non-radiative decay term.
In this paper, we propose a numerical method that can analyze a spontaneous emission process even in a vacuum.
A two-level atom is modeled based on an equation of motion for dipole oscillation,
which radiates an electromagnetic field and interacts with the radiated electromagnetic field,
which means a consideration of the absorption process of the electromagnetic field.
A dipole exchanges electromagnetic field with another dipole by absorbing the radiated field, which leads to an dipole-dipole interaction.
The method does not require a priori knowledge of the decay rate of the dipole current caused by the spontaneous emission in a vacuum.
A spontaneous emission is purely quantum mechanical phenomenon.
However, a classical oscillating current dipole radiates an electromagnetic field and is decelerated by a radiated field, which, in a sense, constitutes a classical description of quantum phenomena.
This model makes no assumptions about the spontaneous emission rate of an dipole in a vacuum.
Therefore, the spontaneous emission process can be consistently analyzed using this model.
Using this model, we calculate the dipole current decay in a vacuum,
where the calculated value of the spontaneous emission rate agrees well with that of the theory.
Rabi oscillation and the enhancement of the emission rate of a dipole embedded in a PhC cavity are also obtained.
The enhancement of the optical emission rate caused by coupling with the cavity's optical mode is successfully simulated.
The advantage of our model is that it incorporates the absorption of the field as well as the emission automatically.
This means the dipole-dipole interaction is included as a fundamental process of the formulation\cite{Dicke}.
The radiation process with a dipole-dipole interaction can also be calculated using this method.

A numerical method that describes the time evolution of an electromagnetic field with a dipole is explained in Sec.~2.
Using the method, we present the calculated result of spontaneous emission in a vacuum in Sec.~3.
Calculated results for the radiation from two excited dipoles and the effect of the dipole-dipole interaction between multiple dipoles are also described.

\section{Numerical model of spontaneous emission}
\begin{figure}[htbp]
\centering\includegraphics[width=7cm]{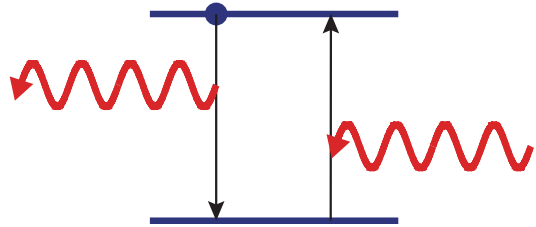}
\caption{Schematic model of two-level system. A dipole in an excited state can transit to a lower state while emitting a photon.  A dipole in a fundamental state can make a transition to an excited state by absorbing a photon.}
\label{fig:m}
\end{figure}

The aim of this study is to analyze the radiation process of a photon from an excited dipole.
In this paper, we assume an atom with two energy levels and model its atomic transition as an oscillating dipole (Fig.~\ref{fig:m}).
An oscillating dipole radiates an electromagnetic field, which can be expressed by the current term in Maxwell's equation.
The time evolution of the electromagnetic field and dipole is expressed by the following Maxwell equation,

\begin{equation}
\nabla\times E(t)=-\mu_0\mu_{\infty}{\frac{d}{dt}}H(t),
\label{eq:eqm1}
\end{equation}
\begin{equation}
\nabla\times H(t)=\varepsilon_0\varepsilon_{\infty}{\frac{d}{dt}}E(t)+j(t),
\label{eq:eqm2}
\end{equation}
where the generation of a magnetic field by an oscillating dipole current is included.
The equation of motion of the dipole current is expressed as
\begin{equation}
{\frac{d^2}{dt^2}}j(t)+\omega_{p}^2j(t)=\varepsilon_0\Delta\varepsilon\omega_p^2{\frac{d}{dt}}E(t),
\label{eq:eqmdp}
\end{equation}
where $\omega_{p}$ is the polarization frequency, and $\Delta\varepsilon=\varepsilon_{s,p}-\varepsilon_{\infty,p}$ is the change in the relative permittivity, where $\varepsilon_{s,p}$ is the static relative permittivity and $\varepsilon_{\infty,p}$is the relative permittivity at infinite frequency.
As can be understood from the equation, $\Delta\varepsilon$ expresses the strength of coupling between dipole and electric field.
The equation describes the time evolution of a dipole accelerated or decelerated by an electric field.
In the equation, $E(t)$ expresses the total electric field radiated by the dipole current and also an external field.
We solve equations (\ref{eq:eqm1}) - (\ref{eq:eqmdp}) using the FDTD method described in \cite{Taniyama11}.

Momentum conservation implies that the particle recoils when it radiates an electromagnetic field \cite{ilderton13}.
This effect is called the radiation reaction, and it is consistently included in Eq.~(\ref{eq:eqmdp}).
In other words, a two-level system in an excited state makes a transition to a fundamental state by emitting a photon with a certain probability.
A transition from a fundamental state to an excited state also occurs when a photon is absorbed.
This could occur even in a vacuum through the re-absorption of an emitted photon.
Both processes are included in Eq.~(\ref{eq:eqmdp}).
When we calculate the time evolution of an initially excited dipole located in a vacuum it emits an electromagnetic field.
The radiated field accelerates or decelerates the motion of the dipole.
Then the amplitude of the dipole current oscillation decays over time.
This can be considered as a classical model of the spontaneous emission process.
As can be understood from Eq.~(\ref{eq:eqmdp}), we assume no artificial damping term of the dipole in our formulation.
An amplitude of the radiated field is proportional to the dipole current as $\sim dj(t)/dt$.
It means the time differentiation of the dipole current term is implicitly included in $E(t)$ of Eq.~(\ref{eq:eqmdp}).
If we express this term explicitly, we obtain the following equation
\begin{equation}
\omega_{p}^2j(t)+2\delta{\frac{d}{dt}}j(t)+{\frac{d^2}{dt^2}}j(t)=\varepsilon_0\Delta\varepsilon\omega_p^2{\frac{d}{dt}}E_{ex}(t),
\label{eq:eqmld}
\end{equation}
where the second term expresses friction and $E_{ex}(t)$ on the right hand side expresses the external field.
In the equation, $2\delta$ is a proportionality constant.
Eq.~(\ref{eq:eqmld}) has the same form as that of Lorentz-dispersive media,
with the optical susceptibility given as
\begin{equation}
\chi(\omega)={\frac{\Delta\varepsilon\omega_p^2}{\omega_p^2-\omega^2+2i\omega\delta}}.
\label{eq:SFDTD}
\end{equation}
In the standard FDTD method for Lorentz-dispersive media, the friction term has been thought to correspond to a non-radiative recombination process \cite{Taflove}.

\section{Spontaneous emission in a vacuum}
Spontaneous emission is a fundamental and purely quantum mechanical phenomenon.
Its exact theoretical description can be achieved using a quantum mechanical expression.
However, classical theory can also provide approximate results for spontaneous emission.
The classical description is based on a classical antenna model.
It describes a quantum state transition caused by an oscillating dipole current.
The model reasonably describes electromagnetic field radiation by a dipole.
The only difference from quantum mechanical theory is that the classical theory gives the half values of the spontaneous emission rate \cite{Shiff,book}.
The fact that a quantum mechanical analysis gives twice the value is attributed to the effect of vacuum fluctuation.
In classical analysis, the vacuum fluctuation is ignored.
Using the expression of optical susceptibility,
\begin{equation}
\chi(\omega)={\frac{N|d|^2}{\varepsilon_0\hbar V}}{\frac{1}{\omega_p-\omega+i\delta}}
\label{eq:SD}
\end{equation}
and Eq.~(\ref{eq:SFDTD}), we obtain the following relation,
\begin{equation}
|d|^2={\frac{1}{2}}\Delta\varepsilon\omega_p\varepsilon_0\hbar V.
\end{equation}
In quantum mechanical theory, the spontaneous emission rate is given as
\begin{equation}
\delta={\frac{\omega^3n}{3\pi\hbar\varepsilon_0c^3}}|d|^2,
\end{equation}
where $d$ is the transition dipole moment.
From this relation, the spontaneous emission rate is expressed using FDTD parameters as
\begin{equation}
\delta={\frac{\omega_p^4n\Delta\varepsilon}{6\pi c^3}}V.
\end{equation}
As a result, we obtain a classical expression of a spontaneous emission rate,
\begin{equation}
\delta={\frac{\omega_p^4n\Delta\varepsilon}{12\pi c^3}}V.
\label{eq:gamma}
\end{equation}

\begin{figure}[htbp]
\centering\includegraphics[width=10cm]{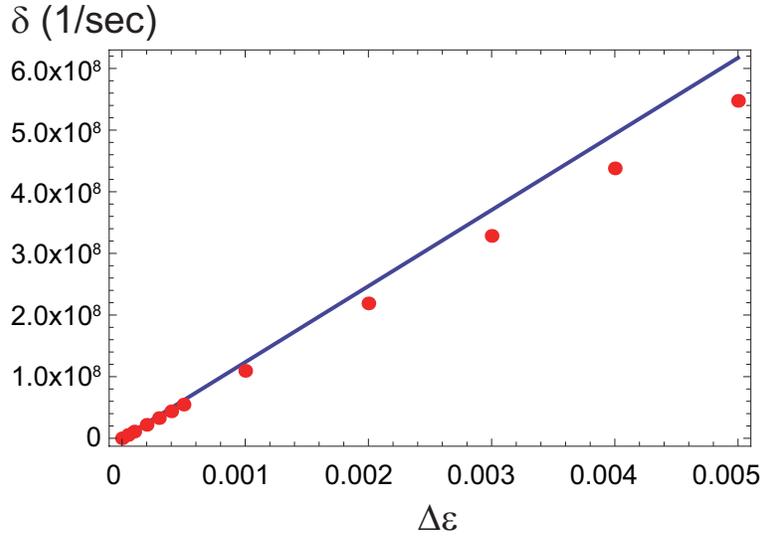}
\caption{Decay time of spontaneous emission process in a vacuum as a function of $\Delta\varepsilon$. The polarization frequency of of the dipole $\omega_p=1.1986\times 10^{15}$Hz. The solid line indicates the theoretical result, and the solid circles are the results of the FDTD calculation.}
\label{fig:s}
\end{figure}

To check the validity of our proposed method, we simulate the radiation process from a dipole located in a vacuum.
As an initial condition of the calculation, we set the dipole in an excited state.
There is no electromagnetic field at a beginning of the calculation.
The dipole radiates an electromagnetic field according to the Eq.~(\ref{eq:eqm2}).
Without right side of Eq.~(\ref{eq:eqmdp}), a dipole oscillation doesn't decay. 
The right side of Eq.~(\ref{eq:eqmdp}) causes the dipole oscillation to be accelerated or decelerated by the electromagnetic field.
A deceleration force make a dipole oscillation decay, which can be understood as recoil force of emission process.
If the dipole feels no recoil force, its amplitude does not decay during the simulation.
However, the simple exponential decay of the dipole current amplitude occurs, which means the recoil force of emission is included in the simulation.
The decay time is estimated by the exponential fitting the time evolution of the dipole current amplitude.
The calculated decay time of the dipole current is shown in Fig.~\ref{fig:s}.
Theoretical values obtained with Eq.~(\ref{eq:gamma}) are also shown in the figure.
As seen in the figure, the results agree well.
Our proposed numerical model can well describe the spontaneous emission process in a vacuum.

\section{Spontaneous emission in a PhC cavity}
An advantage of our model that we wish to emphasize is that it can naturally incorporate a dipole-dipole interaction through a radiation field.
In the proposed numerical model, dipole oscillation can absorb as well as radiate a field.
If there are multiple dipoles, a field radiated by a dipole can be absorbed by another dipole, and vice versa.
This means an interaction between dipoles,
which in turn means that a dipole-dipole interaction is naturally included in the formulation.

\subsection{Spontaneous emission of two excited dipoles}
Planar PhC cavities are attracting growing interest due to their potential for confining light in extremely small volumes and their high $Q$ factor.
The density of state for a photon is greatly modified in a PhC cavity.
The probability of transition from an excited state to the fundamental state is dependent on its environment,
namely the photon density of states.
This is called the Purcell effect and it suggests the possibility of controlling the amount of optical nonlinear effect and also the probability of radiation from excited dipoles artificially.

\begin{figure}[htbp]
\centering\includegraphics[width=16cm]{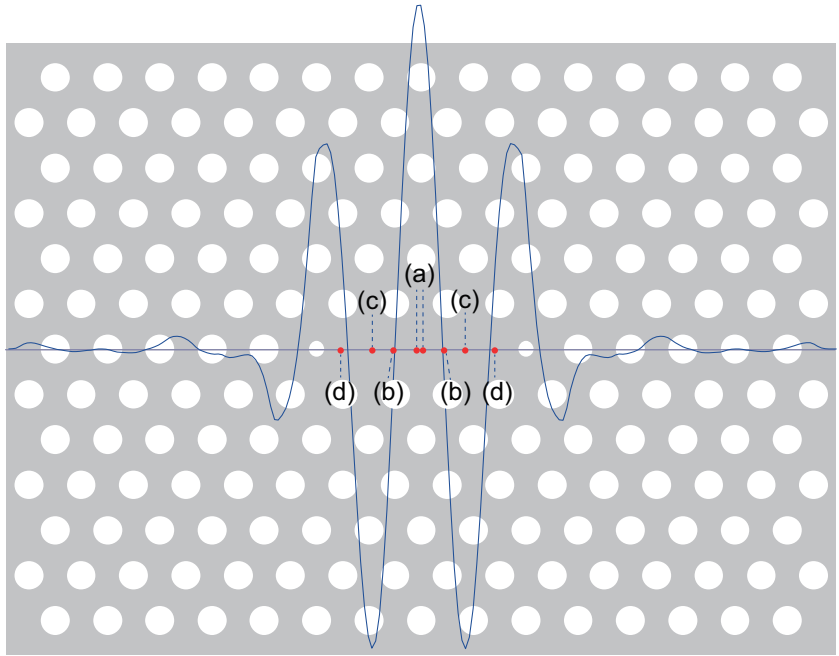}
\caption{Silicon PhC cavity structure with three missing holes (L3). The electric field distribution of cavity's fundamental mode through the center of the cavity is shown. Several configurations of two dipoles in the PhC cavity are studied as indicated in the figure by (a), (b), (c), and (d). The $Q$-factor of the cavity is $1.1\times 10^4$ and the effective mode volume is $0.071\mu m^3$.}
\label{fig:f}
\end{figure}

As an example of a PhC cavity, a PhC cavity with three hole missing (L3) is assumed in the calculation.
The thickness of the PhC slab is $t=210$ nm, the lattice constant is $a=420$ nm,
and the radius of the air hole is $r=115.5$ nm, and the refractive index of the slab material is assumed to be 3.475.
For the fundamental mode, the amplitude of the electric field distribution is shown in Fig.~\ref{fig:f} along with the missing holes of the cavity.
The electric field has an even symmetry and the magnetic field has an odd symmetry in this direction.
The electric field has its maximum value at the center of the cavity and the magnetic field has a node there.
We locate one or two dipoles in the cavity at the center of cavity structure and the middle of slab.
The two dipoles are located one numerical grid (30 nm) apart.
The eigenfrequencies of the dipoles are assumed to be the same.
When there are two dipoles, we assume that both are in excited states with the same phase as the initial condition.
No electromagnetic field is assumed to exist in the cavity at the beginning of the calculation.

\begin{figure}[htbp]
\centering\includegraphics[width=10cm]{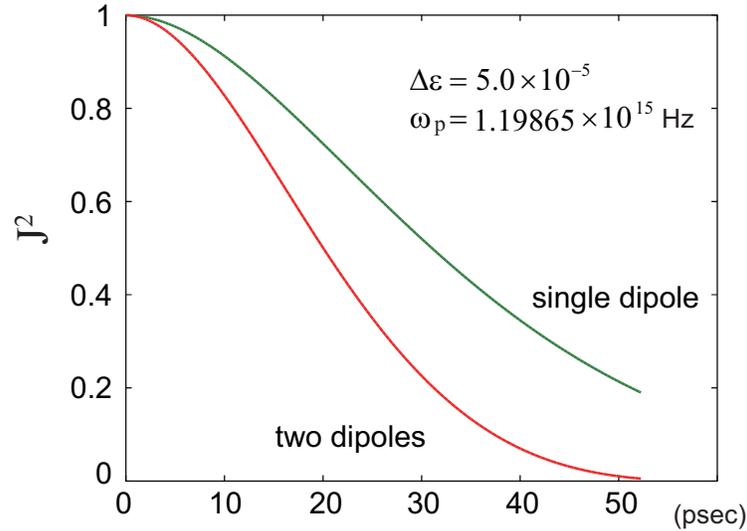}
\caption{Time evolution of dipole current amplitude for single dipole and two dipoles.  The location of the dipoles are the center of a cavity as indicated (a) in Fig.~\ref{fig:f}. In case of two dipoles, they are located one grid (30 nm) apart. The polarization frequency $\omega_p$ is $1.19865\times 10^{15}$Hz and $\Delta\varepsilon=5.0\times 10^{-5}$.}
\label{fig:d}
\end{figure}

Figure \ref{fig:d} shows the dipole current intensity as a function of time for $\Delta\varepsilon=5.0\times 10^{-5}$ and $\omega_p=1.19865\times 10^{15}$ Hz.
At this condition, dipole's eigen frequency corresponds to resonance condition.
The results for one and two dipoles are shown in the figure.
As shown in the previous section, the radiated field affects the dipole radiation process.
Because of Purcell's effect the cavity mode strongly affects the dipole radiation process in a cavity.
The time dependence of the dipole current does not exhibit simple exponential behavior.
This is because the formation of a cavity mode by a radiated field needs a certain amount of time.
In other words, a radiated field does not formulate a cavity mode instantly and gradually affects the radiation process in a cavity.

\begin{figure}[htbp]
\centering\includegraphics[width=12cm]{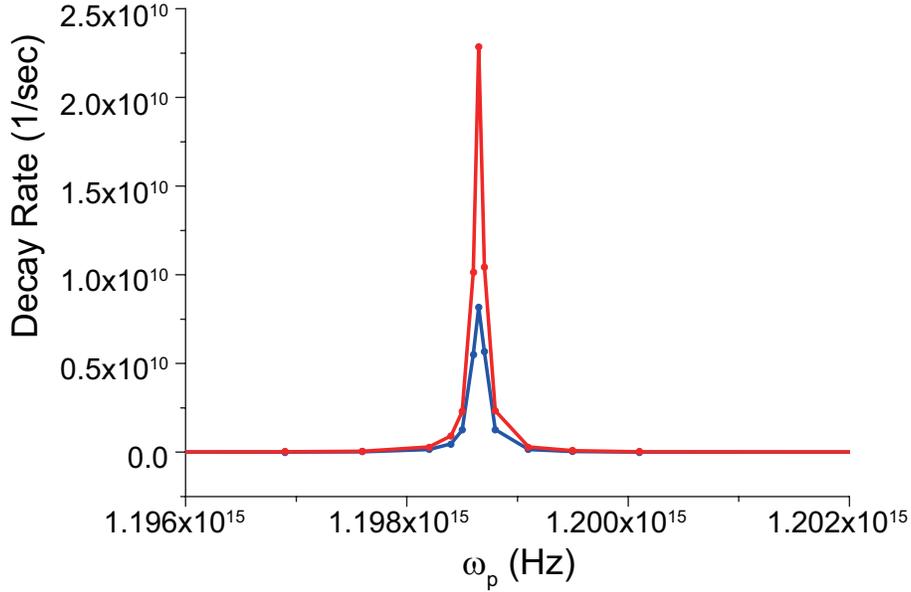}
\caption{Decay rate of excited dipole. Single dipole (blue) and two dipoles (red) are located at the center of the PhC cavity. The polarization frequency $\omega_p=1.19865\times 10^{15}$HZ and $\Delta\varepsilon=5.0\times 10^{-5}$.}
\label{fig:e}
\end{figure}

With two dipoles, the dipole current decays much faster than that of a single dipole, although simple exponential fitting is not possible.
Of course, the decay rate of a dipole strongly depends on the resonant condition of the dipole frequency in the cavity.
To check a resonant condition effect on the radiation process from the dipole, the decay rates of two dipoles are calculated for several eigenfrequency conditions.
The results for both one and two dipoles are shown in Fig.~\ref{fig:e}, which reveals clear resonance in the decay rate.
This indicates a strong enhancement of the radiative decay when the dipole frequency coincides with the cavity's resonance.
For two dipoles, the decay rate in a resonant condition is about twice that of a single dipole.
We consider that the enhancement is induced by the interaction between two dipoles.
We considered that a phenomenon similar to superradiance\cite{Dicke} occurs in this case, although the simulation is purely classical.
Dicke's superradiance process consists of a two-step process.
First, the phases are substantially aligned by the formation of maximally-entangled state through the spontaneous emission, and then the light-emitting body having in-phase acts like a phased array.
Here our model deals with the latter of them.

In the previous calculation, two dipoles are located at the center of cavity with one numerical grid (30 nm) apart, where the electric field of the fundamental mode has its maximum value and their initial phases coincide.
The radiated field and dipole oscillation always positively interfere in this condition.
This interference depends on the relative positions and also on the cavity mode.
To check the dipole location dependence on the radiation process, we performed numerical calculations for different dipole conditions as shown in Fig.~\ref{fig:f}.
The results are shown in table \ref{tab:p}.
(a) are calculations performed for the dipole locations at the center of the PhC cavity, where the fundamental mode of an electric field has it maximum value,
(b) correspond to its node, (c) are at its second maximum points, and (d) are around the middle of the electric field amplitude.
In each condition, two dipoles are located symmetrically opposite in terms of the cavity center and in the same initial phase.
The result indicates that the decay rate of a dipole becomes large at the position where cavity mode has a large amplitude.
In contrast, the enhancement of the decay rate for dipoles at the node position is very small.

\begin{table}[htb]
\centering\caption{The position dependence of the decay rate is shown for two dipoles. The dipole locations are indicated in Fig.~\ref{fig:f}. The polarization frequency is $\omega_p=1.1996\times 10^{15}$Hz and $\Delta\varepsilon=0.003$.}
\begin{tabular}{cp{1.7in}rp{1.7in}}
\hline~
Dipole location & Decay rate (1/sec) \\ \hline
(a) & $2.1367\times 10^{11}$ \\
(b) & $0.1096\times 10^{11}$ \\
(c) & $1.7572\times 10^{11}$ \\
(d) & $0.5797\times 10^{11}$ \\ \hline
\end{tabular}
\label{tab:p}
\end{table}

\begin{figure}[htbp]
\centering\includegraphics[]{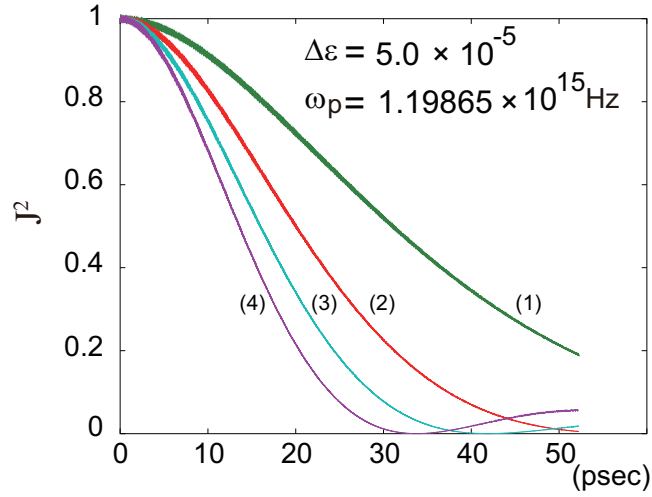}
\caption{Time decay of amplitude of dipole current for (1) single dipole, (2) two dipoles, (3) three dipoles, and (4) four dipoles. The dipoles are located at the center of a PhC cavity and excited in the same phase initially. The polarization frequency is $\omega_p=1.19865\times 10^{15}$Hz and $\Delta\varepsilon=5.0\times 10^{-5}$.}
\label{fig:1234}
\end{figure}

\begin{figure}[htbp]
\centering\includegraphics[width=10cm]{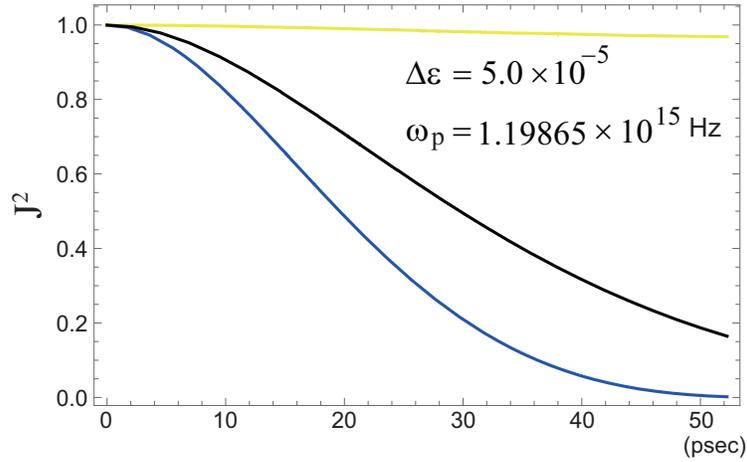}
\caption{Time dependence of dipole current amplitude for (black) single dipole, (blue) two dipoles initially excited in the same phase, and (yellow) two dipoles initially excited in the phase with $\pi$ difference. The polarization frequency is $\omega_p=1.19865\times 10^{15}$Hz and $\Delta\varepsilon=5.0\times 10^{-5}$.}
\label{fig:subrad}
\end{figure}

\subsection{Effect of multiple dipole interaction in the emission process in a PhC cavity}
The cooperative spontaneous emission from a collection of dipoles was studied by Dicke in 1954\cite{Dicke}.
When the dipoles are coherently radiating in phase, the dipoles radiate $N$ times faster than in an incoherent emission, where $N$ is the number of dipoles.
This phenomenon is known as superradiance.
Superradiance in Dicke's theory is quantum mechanical, but classical theory gives an approximate description of the phenomenon.
To check if the proposed method can simulate this phenomenon, we calculate the time evolution of initially excited dipoles in phase.
The dipoles are located in the center of the PhC cavity.
The time evolution of the dipole current is calculated and the time dependence of the amplitude decay is shown in Fig.~\ref{fig:1234}.
As expected, the calculated results show an enhancement in the radiation rate from the dipole.
The amount of enhancement is larger for a larger number of dipoles.
Dicke's work indicated destructive interference as well as constructive interference.
The spontaneous emission of a dipole can be seriously suppressed by destructive interference.
This phenomenon is called subradiance.
To simulate the subradiance that occurs as a result of destructive interference,
we simulate the radiation process of two dipoles located in the center of a cavity.
As an initial condition, the out-of-phase excitation of two dipoles is assumed and compared with two dipoles in-phase excitation and with single dipole excitation.
The result is shown in Fig.~\ref{fig:subrad}.
The figure shows that the decay rate is greatly suppressed with out-of-phase excitation and agrees well with the prediction for subradiance.
As a result, we can say that the model can approximately and reasonably describe the dipole-dipole interaction of radiation processes.

\section{Conclusion}
In summary, we have presented a numerical technique that allows us to analyze fundamental optical processes,
such as the spontaneous emission process of an atom in a vacuum and atoms located in a cavity.
The multi-dipole interaction through a radiated electromagnetic field is included in the method.
The influence of a dipole-dipole interaction on the emission process in a cavity is simulated and both an enhancement and a suppression of the radiation rate are exhibited.
Although the model is based on classical theory, it helps us to better understand the phenomenon.
We believe that the numerical method proposed here would be useful for a better understanding of fundamental physical phenomena such as spontaneous emission processes.
We thank A. Shinya and H. Gotoh for their encouragement throughout this work.

\end{document}